\documentclass[12pt, executivepaper]{article}
\usepackage{times}
\usepackage[utf8]{inputenc}
\usepackage[T1]{fontenc}
\usepackage{url}
\usepackage{amsmath, tabularray}
\usepackage{tabularray}

\usepackage{xcolor}
\usepackage{color, hyperref}
\usepackage{breakcites}
\usepackage{longtable}
\usepackage{fancyvrb}
\usepackage{lipsum}
\usepackage{pbox}

\DefTblrTemplate{middlehead,lasthead}{default}{Continued from previous page}

\newcommand*\justify{%
  \edef\restore{%
    \fontdimen2\font=\the\fontdimen2\font
    \fontdimen3\font=\the\fontdimen3\font
    \fontdimen4\font=\the\fontdimen4\font
    \fontdimen7\font=\the\fontdimen7\font
    \hyphenchar\font=\the\hyphenchar\font
  }%
  \fontdimen2\font=0.5em 
  \fontdimen3\font=0.22222em 
  \fontdimen4\font=0.11111em 
  \fontdimen7\font=0.11111em 
  \hyphenchar\font=`\- 
}

\definecolor{bluer}{RGB}{55, 126, 184}
\definecolor{red}{RGB}{228, 26, 28}
\usepackage{amsfonts}
\usepackage{alltt}
\usepackage{makecell}
\usepackage{graphicx}
\usepackage{tabularray}

\usepackage{enumitem}
\usepackage{geometry}
\usepackage{mathtools}
\usepackage{cellspace, tabularx}
\usepackage{dcolumn, braket, graphicx, xcolor, bm, lipsum, cases, epigraph, tabularx}
\usepackage{hyperref}
\hypersetup{colorlinks= true, allcolors=blue}
\renewcommand{\texttt}[1]{%
  \begingroup
  \ttfamily
  \begingroup\lccode`~=`/\lowercase{\endgroup\def~}{/\discretionary{}{}{}}%
  \begingroup\lccode`~=`[\lowercase{\endgroup\def~}{[\discretionary{}{}{}}%
  \begingroup\lccode`~=`.\lowercase{\endgroup\def~}{.\discretionary{}{}{}}%
  \catcode`/=\active\catcode`[=\active\catcode`.=\active
  \scantokens{#1\noexpand}%
  \endgroup
}

\title{\textbf{Can Machines Philosophize?}}

\bigskip
\author{Michele Pizzochero\thanks{E-mail: mp2834@bath.ac.uk} \\ \emph{\small{Department of Physics, University of Bath}},  \emph{\small{Bath BA2 7AY, United Kingdom}}  \\ \emph{\small{School of Engineering and Applied Sciences, Harvard University}},   \emph{\small{Cambridge, MA 02138, United States}} 
\and  Giorgia Dellaferrera \\ \emph{\small{McKinsey \& Company}}, \emph{\small{London WC1A 1PB, United Kingdom}}}

\begin{document}
\date{}
\maketitle 

\newpage
\begin{abstract} Inspired by the Turing test, we present a novel methodological framework to assess the extent to which a population of machines mirrors the philosophical views of a population of humans. The framework consists of three steps: (i) instructing machines to impersonate each human in the population, reflecting their backgrounds and beliefs, (ii) administering a questionnaire covering various philosophical positions to both humans and machines, and (iii) statistically analyzing the resulting responses. We apply this methodology to the debate on scientific realism, a long-standing philosophical inquiry exploring the relationship between science and reality. By considering the outcome of a survey of over 500 human participants, including both physicists and philosophers of science, we generate their machine personas using an artificial intelligence engine based on a large language model. We reveal that the philosophical views of a population of machines are, on average, similar to those endorsed by a population of humans, irrespective of whether they are physicists or philosophers of science. As compared to humans, however, machines exhibit a weaker inclination toward scientific realism and a stronger coherence in their philosophical positions. Given the observed similarities between the populations of humans and machines, this methodological framework may offer unprecedented opportunities for advancing research in the \textcolor{black}{empirical social sciences} by \textcolor{black}{complementing human participants with their machine-impersonated counterparts.}

\noindent
 \end{abstract} 

\newpage

\small{
\setlength{\epigraphwidth}{0.35\textwidth}
\epigraph{{We are programmed just to do} \\
Anything you want us to\\
We are the robots \\
We are the robots
}{Kraftwerk, \emph{The Robots} (1978)}
}

\section{Machines: Like or unlike humans?}  \textcolor{black}{The question of whether and to what extent machines are akin to humans bears implications for a wide spectrum of disciplines, including the philosophy of mind, the nature of consciousness, intelligence and language, as well as computer science \cite{French:2000}.  The canonical beginning of the discussion is credited to Alan Turing who, in  his now-classic 1950 essay,  posed the question ``Can machines think?'' \cite{Turing:1950}. Turing argues that, formulated as such, the question is ``too meaningless'' and thus advocates for  its replacement with an actionable protocol ``which is closely related to it:''  the  Turing test or imitation game. In the Turing test, a human judge ought to identify, through a text-only conversation, which of the two interlocutors is the machine and which is the human. If the judge fails to reliably distinguish the nature of the interlocutors, then the machine is said to have passed the test, leading to the highly controversial conclusion that it can exhibit an intelligent or human-like behavior. A number of arguments have been leveled against the Turing test as a genuine measure of human intelligence, such as the Chinese Room \cite{Searle:1980} and the Blockhead arguments \cite{Block:1981}, claiming that even non-intelligent systems could potentially pass it. This suggests that inner psychology cannot be merely reduced to the observation of the outward behavior of an agent. Some criticisms have further stimulated the formulation of revised versions of the test, e.g., the inverted \cite{Watt:1996}, questioning \cite{Damassino:2020}, and total Turing tests \cite{Harnad:1991}.} 

\textcolor{black}{The development of approaches to detect the presence of thought in putatively minded agents traces back to Cartesian philosophy. In the \emph{Discourse on the Method}, Descartes proposes an approach for distinguishing humans from automata, arguing that the latter are invariably incapable of being convincingly disguised as the former. Gunderson has noted that Descartes in fact distinguishes two tests \cite{Gunderson1964}. First, the language test, whereby a machine ``could never use speech or other signs placing our thoughts on record for the benefit of others,'' for
``it never happens that it arranges its speech in various way, in order to reply appropriately to \textcolor{black}{everything that may be said in its presence,} as even the lowest type of man can do.'' Second, the action test, which is based on the idea that  ``although machines can perform certain things as well as or perhaps better than any of us can do, they infallibly fall short in others, by the which means we may discover that they did not act from knowledge, but only from the disposition of their organs. For while reason is a universal instrument which can serve for all contingencies, these organs have need of some special adaptation for every particular action. From this it follows that it is morally  impossible that there should be sufficient diversity in any machine to allow it to act in all events of life in the same way as our reason causes us to act'' \cite{Gunderson1964}. Erion has offered an alternative interpretation of the action test by arguing that it has to be regarded  as a test of
common sense, understood as the ability to perform tasks that even the most simple-minded adult human can do \cite{Erion2001}. Be that as it may, Turing was aware of the Cartesian language test, which played an important role in the introduction of the imitation game \cite{Abramson2011}.}

\textcolor{black}{The exploration of the analogies and differences between humans and machines has recently experienced a new renaissance following the rapid progress of generative Artificial Intelligence (AI) systems \cite{Mitchell:2024} such as large language models \cite{Naveed:2024}---a class of neural networks trained on vast amounts of text data to understand, analyze, and generate natural language---along with their impressive contribution to science \cite{Krenn:2022, Birhane2023} and society \cite{Weidinger:2022}. Large language models appear to instantiate what Descartes in the  \emph{Discourse}  deemed  ``not conceivable,'' namely, the realization of ``a machine'' that  ``should produce different arrangements of words so as to give an appropriately meaningful answer to whatever is said in its presence, as the dullest of men can do.''}  To ascertain the ability of AI systems to imitate humans, several investigations have implemented Turing-like tests that diverge from the original proposal. Instead of resorting to iterative, language-based interactions with a machine, such tests evaluate the performance of AI models by directly comparing their outputs against the ground truth\footnote{\textcolor{black}{Here, ‘ground truth’ is used in the statistical sense—not the philosophical one—referring to the ideal expected result against which the performance of a machine is evaluated.}} generated by their human counterparts, without involving the mediation of a human judge. These Turing-like tests have so far been mainly restricted to tasks that do not pose any considerable challenge to humans and that deliver an outcome that can be classified as either correct or incorrect, such as visual question answering \cite{Ming:2023}, image captioning \cite{Kasai:2022} and recognition \cite{Dodge:2017}. However, the capability of current AI models to emulate humans in activities that involve thinking in complexity has hitherto remained largely unexplored. Addressing this issue is particularly timely, in that the artificial intelligence encoded in large language generative models, as quantified by the number of computational parameters, has approached the biological intelligence encoded in \emph{homo sapiens}, as quantified by the number of synapses in the brain  \cite{Schwartz:2022}. 

\textcolor{black}{Here, we consider the question: ``Can machines philosophize?''} Philosophizing is generally regarded as a uniquely human pursuit, embedded in a collective and multifaceted endeavor, demanding, \textcolor{black}{for example,} internal consistency rather than right-or-wrong classifications, and often shaped by factors that are typically inaccessible to machines.  This latter aspect is corroborated by recent empirical studies indicating that philosophical judgments are influenced by, e.g.,  birth cohort \cite{Hannikainen:2018}, gender \cite{Buckwalter:2013}, alcohol consumption \cite{Duke:2015}, wording  and ordering effects \cite{Petrinovich:1996}, as well as personality traits \cite{Bartels:2011},  as hypothesized by William James the beginning of the twentieth century \cite{James:2000}.\footnote{\textcolor{black}{We do not take a normative stance on whether such influences should or should not occur, as this question is irrelevant the scope of the present work.}} Inspired by the Turing test, we design and implement a novel methodological framework to quantify the degree to which the philosophical positions held by machines mirror those of humans.

 \textcolor{black}{{The rationale of our study is twofold, encompassing both fundamental and applied dimensions. On the fundamental side, previous works comparing humans and machines---typically framed in the tradition of the Turing test---have largely concentrated on the assessment of \emph{individual} agents. By contrast, our approach shifts the focus from individuals to \emph{populations}, asking whether AI can emulate the statistical distribution of views that emerges in human groups. This population-level perspective not only extends the methodological scope of the imitation game but also opens novel questions about the capacity of machines to reproduce patterns of collective human reasoning. On the applied side, our framework may offer practical benefits for quantitative research in the empirical social sciences, where survey-based methods are widely used to chart attitudes and inform debates. In these contexts, machine-generated populations with controlled characteristics could complement, or in some cases partially substitute for, human subject pools, thereby broadening the methodological toolkit available to researchers while helping overcome some challenges  that are inherent to survey-based studies, such as recruiting a sufficiently large and representative sample, high costs and long timelines required to gather responses, or survey design, given that poorly worded questions may only become evident once responses are collected. }}

In this work, first we devise a general protocol to instruct large language, generative AI models to impersonate a \emph{population} of humans, reflecting the diverse backgrounds and beliefs of individuals. Second, we apply this protocol to compare the philosophical positions endorsed by a population of humans with those endorsed by the corresponding population of machines. \textcolor{black}{{As a case study, we consider the debate on scientific realism \cite{Psillos1991}---a century-long inquiry into the relationship between science and reality---owing to its central role in the philosophy of science.  This choice is guided by both philosophical and empirical considerations. From a philosophical perspective, scientific realism stands as one of the most enduring and contested issues in philosophy, engaging the sustained attention of both scientists and philosophers. Its centrality and unresolved status make it an especially informative case for evaluating whether machines are capable of navigating and reproducing reasoning within a live, interdisciplinary controversy. Indeed, unlike many applications where AI outputs are checked against an objective ground truth (e.g., image captioning), issues that emerged within the scientific realism debate typically involve multiple, coexisting perspectives rather than a single, definitive answer. From an empirical perspective, our study builds upon the work of Henne and coworkers \cite{Henne2024}, which offers one of the most recent and comprehensive surveys available on attitudes toward scientific realism. This survey not only provides a robust benchmark for evaluating AI-generated responses but also encompasses a heterogeneous population, including physicists from a range of subfields and philosophers subscribing to diverse positions. Such diversity allows us to examine whether machines can approximate not only the aggregate tendencies of human respondents but also the more fine-grained perspectives of different disciplinary groups.}}

Our findings indicate that a population of machines holds beliefs that are, on average, quite similar to those held by the population of humans, differing only by a few percent.  As compared to humans, however, machines exhibit a slightly less pronounced inclination toward scientific realism and significantly more coherent philosophical views. We additionally identify common patterns underlying human and machine populations when confronted with the philosophical challenges raised by the realism debate. Overall, our analysis unveils the ability of machines to imitate human populations when addressing complex issues, possibly paving the way for the introduction of AI-assisted approaches in the \textcolor{black}{empirical social sciences.}

\section{Philosophical views in the scientific realism debate}
 Our work builds on the approach of Henne and coworkers \cite{Henne2024}, which employs a questionnaire to survey the views of physicists and philosophers of science within the scientific realism debate. The questionnaire consists of 30 statements listed in Table \ref{Table-1} describing, either directly or indirectly through specific examples, four philosophical positions, that is, scientific realism (including several forms of selective realism), instrumentalism, pluralism, and perspectivism. We briefly outline these positions.

{Scientific realism} \cite{Psillos1991} is the view that science portrays a faithful representation of the world (S1 and S2) by discovering objects that are beyond our perception (S4 and S6), such as electrons (S13, S15, S17, S18), and formulating theories that are at least approximately true (S8), such as the Big Bang theory, as an example of speculative physics (S20),\footnote{{\textcolor{black}{Speculative physics involves theoretical frameworks and conjectures to explore phenomena beyond currently available empirical evidence and testability.}}} or general relativity (S21).  This view is rooted in three commitments. First, a metaphysical commitment, holding that there exists a mind-independent reality, namely, a reality that is not sensitive to our specific theories (S14) or our particular approach to manipulate and describe it (S11).  Second, a semantic commitment, holding that successful scientific theories ought to be interpreted as (approximately) true by correspondence (S21), thus revealing the actual features of reality (S23), e.g., the nature of space and time (S22). \textcolor{black}{Third, an epistemic commitment, holding that belief in scientific theories is justified by compelling epistemic reasons}.  \textcolor{black}{These commitments are often complemented by the axiological claim that the scope of science is to achieve truth by correspondence (S10).}

\textcolor{black}{A prevailing variety of scientific realism is selective realism \cite{Chakravartty2010}, which prescribes that that the ontological commitment should not be endowed to scientific theories \emph{en bloc}.} Instead, belief should be restricted to  a narrow subset of theoretical claims. Depending on the subset of theoretical claims regarded as epistemically secure, three main forms of selective realism have been developed, that is, deployment realism \cite{Psillos1991},  structural realism, and entity realism. Structural realism \cite{Worrall1989, Ladyman1998} warrants belief in relations, as typically subsumed in the mathematical structures of scientific theories, while advocating for skepticism about entities (S27). Entity realism \cite{Hacking1983, Cartwright1983}  warrants belief in entities, especially those that are amenable to experimental manipulation, while advocating for skepticism about the mathematical structures that purport to describe them (S13, S14, S18, S19).

\textcolor{black}{One of the views opposing scientific realism is {instrumentalism}} \cite{Rowbottom2019, Stanford2010},\footnote{\textcolor{black}{Other relevant antirealist views are, e.g., positivism and, more recently, constructive empiricism \cite{vanFraassen:1980}}}  which denies that successful scientific theories can offer access to the ultimate nature of reality (S3). Rather, it acknowledges the effectiveness  of scientific theories as devices to classify and predict phenomena (S9, S23), while considering their posited unobservable entities being mere fictions (S5) or constructions (S7). Instrumentalism is often driven by a `pessimistic meta-induction' \cite{Laudan1981} which draws from the history of science to infer that present-day theories, analogous to the superseded theories of the past, are likewise poised to be abandoned (S12). Unlike scientific realism, instrumentalism rejects the notion that scientific theories are conducive to truth by correspondence, regarding the unobservable entities existing only within the sole province of the theories that postulate them.

Besides the realism-instrumentalism dichotomy, the debate has given rise to additional positions, such as pluralism and perspectivism. \textcolor{black}{{Pluralism} \cite{Chang:2012, Massimi:2018} has been articulated in many flavors, some compatible with some forms of realism, other not.}  \textcolor{black}{Epistemic pluralism, for instance, maintains that multiple and even incompatible theories can nevertheless be of equal epistemic value, while ontological pluralism embraces the idea that the nature of the world is not unique (S24, S25).} \textcolor{black}{Methodological pluralism maintains that conflicting theories are valuable for the progress of scientific inquiry (S28).} {Perspectivism} \cite{Massimi:2022, Giere:2006, Lipton:2007} advocates that different theories entail different perspectives on the world, each of them delivering a representation from a particular and limited point of view (S30) that is influenced by the cultural traditions and historical periods in which the theories are formulated (S29).  \textcolor{black}{{Closely related to these position is internal realism \cite{Putnam:1981} which rejects the ‘God’s-Eye Point of View’ inherent in metaphysical realism---the view that the external reality is objective and independent of how inquiring agents conceptualizes it---holding instead that our understanding and the world and the truth of our theories are relative to a given conceptual scheme (S16). }}

\begin{longtblr}[
  caption = {List of statements assessed by physicists, philosophers of science, and their corresponding machine-generated personas, covering various philosophical positions that emerged in the scientific realism debate.  \textcolor{black}{Statements highlighted in blue (red) denote agreement (disagreement) with scientific realism,} whereas statements highlighted in grey may be compatible with both views and include pluralism and perspectivism.} \textcolor{black}{Reproduced verbatim from the survey of Henne and coworkers \cite{Henne2024}. },
  label = {Table-1},
]{
 colspec={|X[1]|X[18]|X[8]|},
  rowhead = 1,
  hlines,
} 
\centering \textbf{S}       &  \centering \textbf{Statement} & \centering \textbf{Philosophical position} \\
\centering \textcolor{bluer}{1}  &    Our most successful physics shows us what the world is really like.  &  \centering Scientific realism (strong)  \\
\centering \textcolor{bluer}{2}  &    Physics uncovers what the universe is made of and how it works.    &    \centering Scientific realism (moderate)\\
\centering \textcolor{black}{3}  &    Our most successful physics is useful in many ways, but physics does not reveal the true nature of the world.  &    \centering Instrumentalism   \\
\centering \textcolor{bluer}{4}  &    The imperceptible objects that are part of our most successful physics probably exist. (with “imperceptible” we mean objects that cannot be perceived with our unaided senses, e.g. electrons, black holes, ...)   &    \centering Entity realism   \\        
\centering  \textcolor{black}{5}  &    The imperceptible objects postulated by physics are only useful fictions.   &    \centering Fictionalism, instrumentalism   \\        
\centering \textcolor{bluer}{6}  &    Physicists discover imperceptible objects.   &    \centering Scientific realism   \\
\centering \textcolor{black}{7}  &    Communities of physicists construct imperceptible objects.  &    \centering Constructivism, instrumentalism   \\ 
\centering \textcolor{bluer}{8}  &    Our best physical theories are true or approximately true.  &    \centering Scientific realism   \\
\centering \textcolor{black}{9}  &    Physical theories do not reveal hidden aspects of nature. Instead, they are instruments for the classification, manipulation and prediction of phenomena.   &   \centering  Instrumentalism   \\   \centering \textcolor{bluer}{10}  &    The most important goal of physics is giving us true theories.   &    \centering Realism about goal   \\
\centering \textcolor{bluer}{11} & If there was a highly advanced civilization in another galaxy, their scientists would discover the existence and properties of many of the imperceptible objects of our current physics.  &    \centering Scientific realism, metaphysical realism.  \\
\centering \textcolor{black}{12}  & I expect the best current theories in physics to be largely refuted in the next centuries---in the same way that successful theories were largely refuted in the past. &    \centering Scientific antirealism, Pessimistic meta-induction  \\
\centering \textcolor{bluer}{13}  & Electrons exist.   &    \centering Entity realism \\
\centering \textcolor{bluer}{14}  & Electrons, with all their properties, exist “out there,” independently from our theories. &    \centering Entity realism, metaphysical realism  \\
\centering \textcolor{bluer}{15}  & Our theories are getting closer to the real nature of the electron. &    \centering Entity realism, metaphysical realism  \\
\centering \textcolor{black}{16}  & Electrons are postulated as real within our models; it does not make sense to ask whether they exist “outside” or independently of the theory/model. &    \centering Internal realism  \\
\centering \textcolor{bluer}{17}  & There is something in the world that behaves like (what we would define as) an electron. &    \centering Entity realism, internal realism  \\
\centering \textcolor{bluer}{18}  & Electrons are (at least) as real as toe-nails and volcanoes. &    \centering Entity realism  \\
\centering \textcolor{gray}{19}  & Phonons exist. &    \centering Entity realism  \\
\centering \textcolor{gray}{20}  & There really was a Big Bang. &    \centering Scientific realism, speculative physics \\
\centering \textcolor{bluer}{21}   & General relativity is a true theory. &    \centering Scientific realism \\
\centering \textcolor{bluer}{22}   & General relativity teaches us about the nature of spacetime. &    \centering Scientific realism \\
\centering \textcolor{black}{23}   & General relativity is not the revelation of an underlying order of nature. It is a tool that helps us make predictions and construct GPS, for example. &    \centering Instrumentalism\\
\centering \textcolor{gray}{24}   & Newtonian mechanics is a true theory. &    \centering Scientific realism, pluralism \\
\centering \textcolor{black}{25}   & If a phenomenon can be explained both by a classical model and by a quantum model, neither of the models is closer to the truth than the other. &    \centering Epistemic pluralism or antirealism \\
\centering \textcolor{gray}{26}   & We should build a particle collider that is bigger than the LHC. &    \centering --- \\
\centering \textcolor{gray}{27}   & A physical theory cannot tell us what the universe is really made of, but the mathematical structure of our best theories represents the structure of the world. &    \centering Structural realism \\
\centering \textcolor{gray}{28}   & Having mutually conflicting theories about the same phenomena is valuable for physics. &    \centering Methodological pluralism \\
\centering \textcolor{gray}{29}   & Our scientific knowledge is the product of the prevailing cultural traditions and historical periods in which they were formulated. &    \centering Cultural/historical perspectivism \\
\centering \textcolor{gray}{30}   & Scientific theories and models are idealized structures that represent the world from particular and limited points of view. &    \centering Perspectivism \\
\end{longtblr}

\section{Assessing the views of  machines}  
\textcolor{black}{Of the 30 statements listed in Table \ref{Table-1}, 14 reflect an inclination toward scientific realism (S1, S2, S4, S6, S8, S10, S11, S13, S14, S15, S17, S18, S20, S22), 8 reflect an opposition to realism }(S3, S5, S7, S9, S12, S16, S23, S25), while the remaining 8 are compatible with both positions (S19, S20, S24, S26, S27, S28, S29, S30). To assess which views, among those presented in Table \ref{Table-1}, are endorsed by physicists and philosophers of science, Henne and coworkers \cite{Henne2024}  administered these statements to 535 participants---consisting of 384 physicists and 151 philosophers of science---requesting them to rate their agreement by assigning an `agreement score' ranging from 0\% (``This sounds completely wrong to me'') to 100\% (``This strikes me as exactly right''). Because participants may not be familiar with the specialized terminology involved in the philosophical debate, they were instructed to assign the agreement score according to their immediate understanding of the statement (``Many of the statements may seem unclear. For example, terms like ‘truth’ and ‘reality’ can be understood in many ways. Please answer according to your immediate inclination'').

\begin{figure}[t]
    \centering
    \includegraphics[width=1\textwidth]{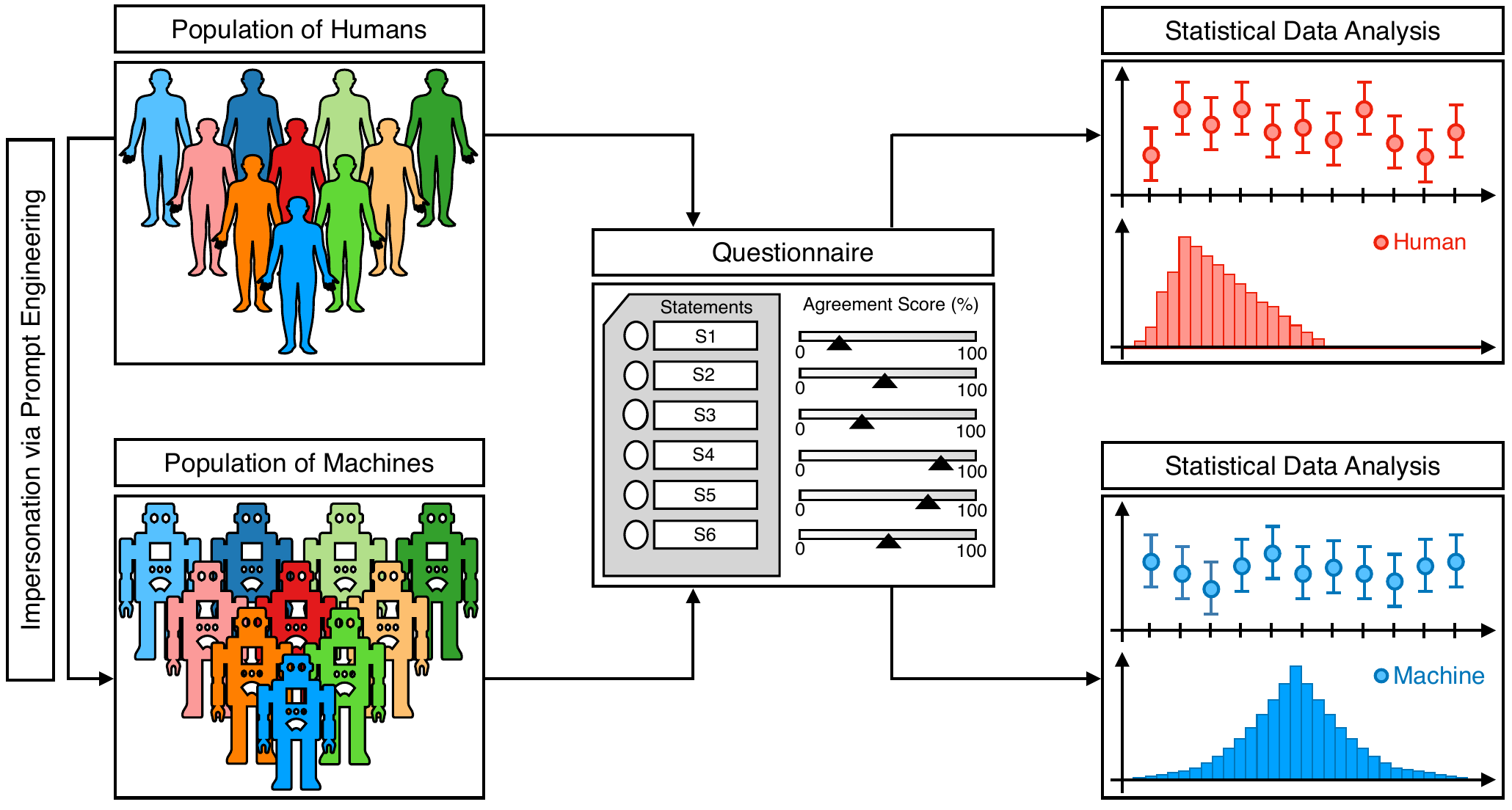}
    \caption{\textbf{Schematic illustration of the methodology developed}. Our methodology unfolds in three steps. First, a population of machines is prompted to impersonate a population of humans, reflecting the academic profile or philosophical beliefs of each individual. Second, each individual in the two populations is administered a questionnaire consisting of a list of statements covering various philosophical positions. Individuals are requested to rate their agreement with each statement by assigning an `agreement score' on a scale ranging from 0\% to 100\%. Third, the results of the survey are statistically analyzed to quantify the analogies and differences in the philosophical views held by a population of humans and the corresponding population of machines.}
     \label{Figure-1}
\end{figure}

For each participant, background information concerning their academic profile was collected. 
For physicists, this included (i) whether their work tends to be theoretical or experimental, (ii) whether their work is rather basic or applied research, (iii) the number of years they have been doing research in physics from the start of their PhD, with available options being 0, 1, 2, 3, 4, 5, 6, 7, 8, 9, 10, 10-15, 15-20, 20-25, or 25+ years, and (iv) what their field of research is, with available options being astrophysics, nuclear and particle physics, atomic, molecular and optical physics, condensed matter physics, applied physics, or other. For philosophers, information included the preferred position within the scientific realism debate, designated by selecting one or multiple options among scientific realism, instrumentalism, constructive empiricism, entity realism, structural realism, perspectivism, pluralism, social constructionism, relativism, logical empiricism, and others to be specified. The data resulting from the survey of Henne and coworkers \cite{Henne2024} is publicly available on Mendeley Data \cite{Sperber:2023}.

 \textcolor{black}{{While the approach of Henne and coworkers \cite{Henne2024}  may carry certain limitations---for instance, allowing participants who may not be fully familiar with the specialized terminology of the scientific realism debate, asking them to rate statements based on their immediate understanding, or an ambiguous formulation of the statements\footnote{\textcolor{black}{Interpretations of the statements listed in Table  \ref{Table-1} that are alternative to those provided by Henne and coworkers \cite{Henne2024} are possible. For example, it is questionable that S1 represents strong realism while S2 weak realism; S1 and S15  do not necessarily entail metaphysical realism; S30 is compatible with perspectivism, but also with standard realism, due to the polysemic nature of the term 'perspectivism.'}}---our aim is not to evaluate these methodological choices. Rather, we apply the framework as established in the literature, since our primary objective is to conduct a rigorous comparison between human and machine responses. To achieve such rigor, it is essential to employ the very same methodology for both populations to ensure that any observed differences reflect the agents themselves rather than variations in the protocols.}}
 
To compare the philosophical views of humans with those of machines, we rely on large language models. Through prompt engineering, we configure this AI engine to impersonate each of the 535 humans participating in the survey of Henne and coworkers \cite{Henne2024}. Our prompt is composed of two parts. In the first part, the AI persona, whether a physicist or philosopher of science, is generated on the basis of the background information collected for the respective human counterpart. In the second part, the resulting AI persona is required to rate their agreement with each of the 30 statements listed in Table \ref{Table-1}, using the exact same formulation of the question that was proposed to the human participants in the survey of Henne and coworkers \cite{Henne2024}. Our methodology is schematically summarized in Figure \ref{Figure-1}. For example, the prompt used to instruct the AI engine to impersonate an experimental physicist conducting applied research, with more than 3 years of experience since the beginning of their PhD, and working in the field of nuclear and particle physics, is as follows:

\begin{quote}
  \par\addvspace{\topsep}
  \begingroup
  \setlength{\parindent}{0pt}%
  \ttfamily
  \justify
You are a physicist. Your work tends to be more experimental. Your work is rather applied research. Since the start of your PhD, you have been doing research in physics for 3 years. Your field of research within physics is nuclear and particle physics. The goal of this survey is to test your reaction towards 30 philosophical statements about physics. Many of the statements may seem unclear. For example, terms like ‘truth’ and ‘reality’ can be understood in many ways. Please answer according to your immediate inclination. Please rate the following 30 statements on a continuous scale between 0 (“This sounds completely wrong to me”) and 100 (“This strikes me as exactly right”). Provide only a single natural number as an answer for each question. Return as output a string containing 30 numbers separated by commas with the format "s1, s2, s3, s4,...,s30" where s1 is the answer to S1, s2 to S2 and so on. Statements: S1: Our most successful physics shows us what the world is really like. S2: Physics uncovers what the universe is made of and how it works. S3: Physics is useful in many ways, but it does not reveal the true nature of the world. S4: [...]
  \par
  \restore
  \endgroup
  \addvspace{\topsep}

\end{quote}
In a similar vein, the prompt used to instruct the AI to impersonate a philosopher of science subscribing to structural realism is as follows:

\begin{quote}
  \par\addvspace{\topsep}
  \begingroup
  \setlength{\parindent}{0pt}%
  \ttfamily
  \justify
  You are a philosopher of science. Your position in the debate on scientific realism is structural realism. The goal of this survey is to test your reaction towards 30 philosophical statements about physics. Many of the statements may seem unclear. For example, terms like ‘truth’ and ‘reality’ can be understood in many ways. Please answer according to your immediate inclination. Please rate the following 30 statements on a continuous scale between 0 (“This sounds completely wrong to me”) and 100 (“This strikes me as exactly right”). Provide only a single natural number as an answer for each question. Return as output a string containing 30 numbers separated by commas with the format "s1, s2, s3, s4,..., s30" where s1 is the answer to S1, s2 to S2 and so on. Statements: S1: Our most successful physics shows us what the world is really like. S2: Physics uncovers what the universe is made of and how it works. S3: Physics is useful in many ways, but it does not reveal the true nature of the world. S4: [...]
  \par
  \restore
  \endgroup
  \addvspace{\topsep}

\end{quote}
\textcolor{black}{To implement our methodology, we developed a Python-based Jupyter Notebook that connects to GPT models through the OpenAI Application Programming Interface (API). Access to the model is managed via a custom function, which handles tasks such as authentication, sending prompts, configuring parameters, and returning structured responses.} The custom function constructs personalized prompts to impersonate each physicist and philosopher of science who participated in the survey of Henne and coworkers \cite{Henne2024}, presents to each resulting AI-generated persona the 30 statements listed in Table \ref{Table-1},  retrieves the agreement scores assigned, and collects them in a dataset. \textcolor{black}{We emphasize that we used GPT-3.5-turbo, which was released prior to the publication of Henne and coworkers  \cite{Henne2024}. Thus, the survey results that serve as our benchmark could not have been part of the  training set of the model. This ensures that the outputs of the model are not simply reproductions of memorized data (i.e., data contamination), but rather the product of more complex generative processes.}

\begin{figure}[t]
    \centering
    \includegraphics[width=1\textwidth]{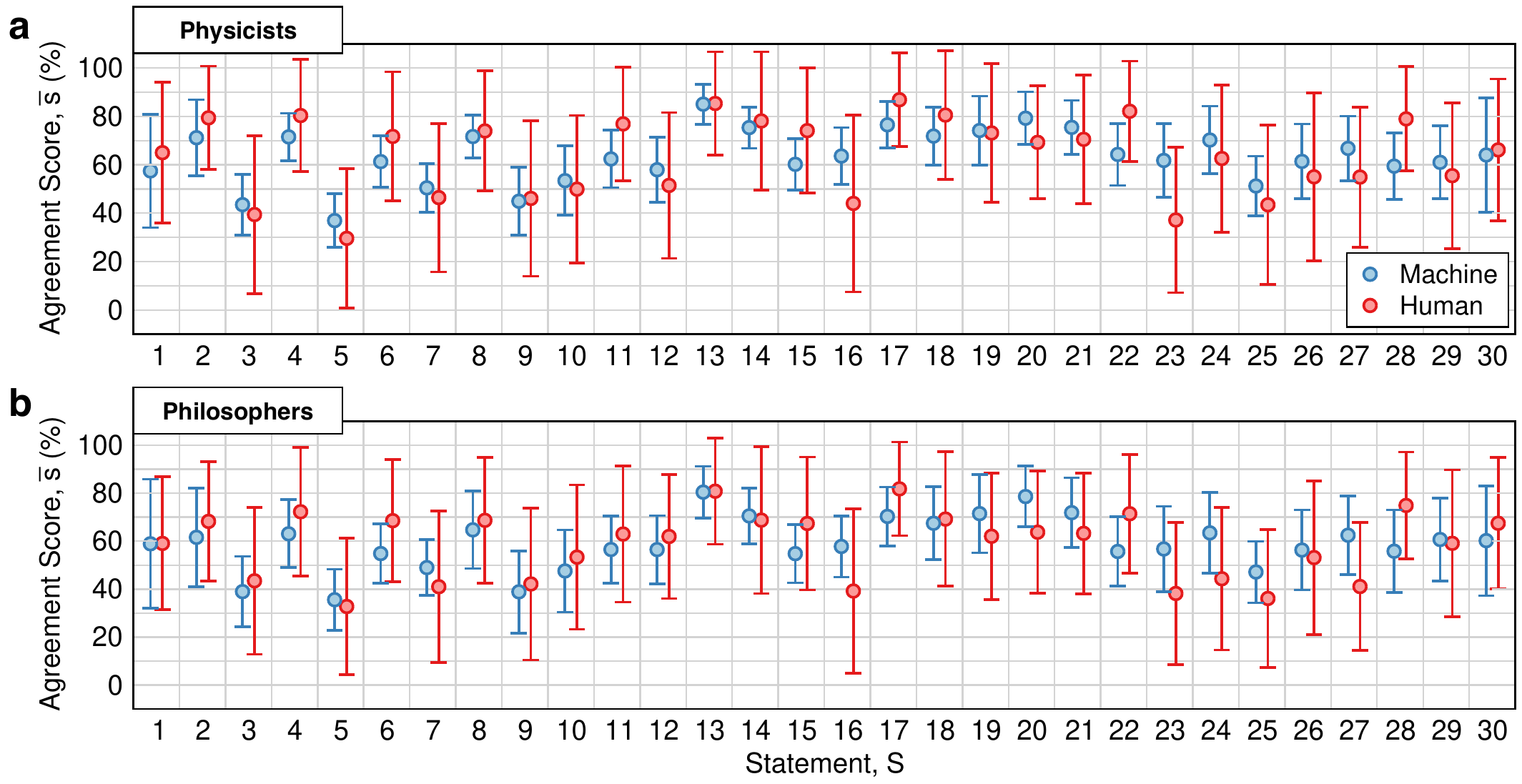}
    \caption{\textbf{Assessment of the philosophical views of humans and machines}. Mean agreement score ($\overline{s}$) assigned to each of the 30 statements ($S$) listed in Table  \ref{Table-1} by a population of machines (blue) and humans (red) consisting of \textbf{a,} physicists and \textbf{b,} philosophers of science.}
     \label{Figure-2}
\end{figure}

\section{Comparing the views of humans and machines} 
We begin the discussion of our results by examining the mean agreement scores of humans and machines for each of the 30 statements listed in Table \ref{Table-1}. The results obtained for the populations of physicists and philosophers of science are shown in Figure \ref{Figure-2}. The corresponding numerical data are provided in Supplementary Table 1. We note that, for each statement, the values of mean agreement scores of machines are comparable to those of humans. Specifically, of the 30 statements administered to physicists, the absolute difference in the mean agreement scores  between humans and machines is less than 10\% for 21 statements and less than 5\% for 10 statements, reaching its minimum of 0.3\% for S13 and its maximum of 24.6\% for S23. Similarly, for philosophers of science, this absolute difference is less than 10\% for 19 statements and less than 5\% for 10 statements, reaching its minimum of 0.2\% for S1 and its maximum of 21.4\% for S27.  The similarities of the mean agreement scores assigned by humans and machines are further highlighted in Supplementary Figure 1 and, as suggested by the additional analysis reported in Supplementary Figure 2, are not affected by any systematic bias. 

\begin{figure}[t]
    \centering
    \includegraphics[width=1\textwidth]{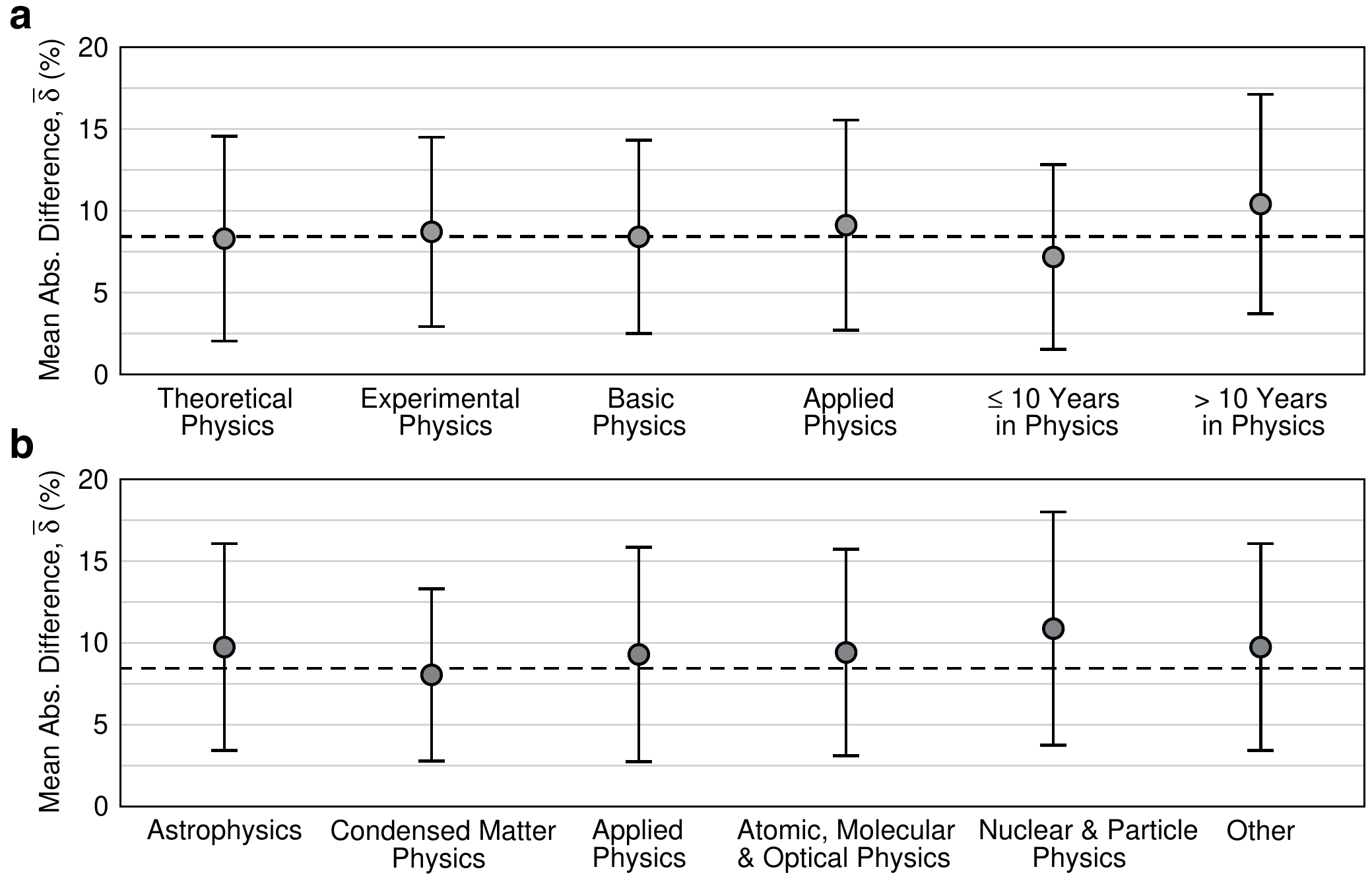}
    \caption{\textbf{Comparison between humans and machines across sub-populations.} Mean absolute difference in mean agreement scores ($\overline{\delta}$) between populations of humans and machines, as defined in Equation \ref{delta}, across different sub-populations of physicists, grouped according to their \textbf{a,} research methodology, scope, or experience, and  \textbf{b,} area of expertise. The dashed horizontal line denotes the global absolute difference obtained for the entire population of physicists, $\overline{\delta}$ = 8.4\%.}
         \label{Figure-3}
\end{figure}

We observe that the standard deviations of the mean agreement scores of humans are consistently larger than those of machines, as illustrated in Figure          \ref{Figure-2} and Supplementary Figure 3, with their difference averaging to 14.7\% and 12.2\% for physicists and philosophers of science, respectively. Importantly, the standard deviation of the mean agreement scores of machines overlaps considerably with those of humans across all statements. This indicates that the philosophical views held by a population of machines resemble very closely those held by a population of human physicists or philosophers of science. We quantify the global discrepancy between humans and machines by determining the mean absolute difference,\footnote{\textcolor{black}{By using the mean \emph{absolute} difference in agreement scores---instead of the \emph{signed} difference---we ensure that our findings cannot be due to error cancellations.}}
\begin{equation}
\overline{\delta} = <\overline{s}\textsubscript{H} > - < \overline{s}\textsubscript{M}>,
\label{delta}
\end{equation}
where $\overline{s}\textsubscript{H}$ and $\overline{s}\textsubscript{M}$ are the mean values of the agreement scores assigned by the human and machine populations, respectively, shown in Figure \ref{Figure-2}. We obtain $\overline{\delta} = 8.4\% \pm 6.0 $ for physicists and $\overline{\delta} = 8.9\% \pm 6.2 $ for philosophers of science. This demonstrates that machines parallel the judgments of both human physicists and philosophers of science equally well, despite their distinct academic profiles.

\begin{figure}[t]
    \centering
    \includegraphics[width=0.8\textwidth]{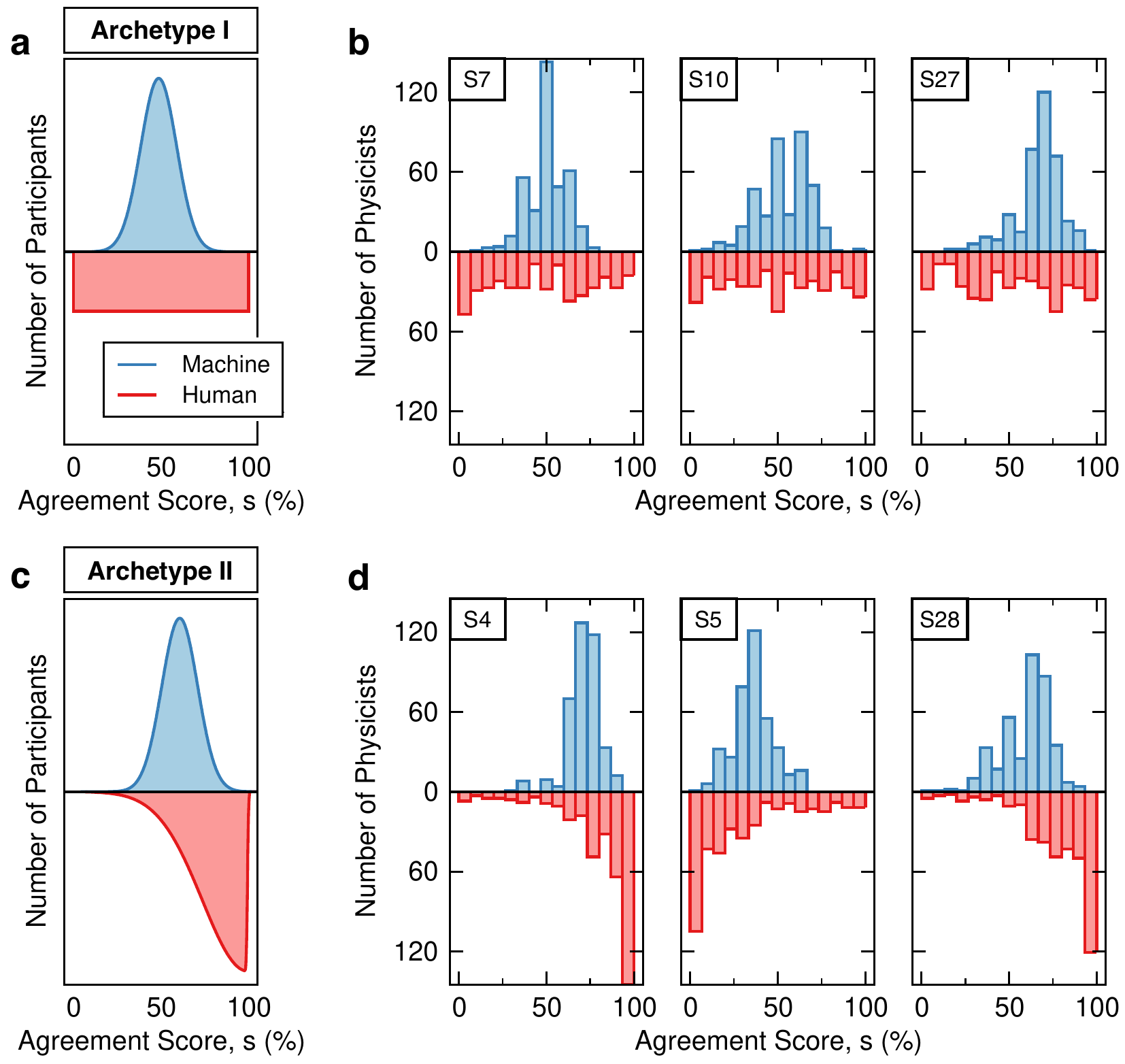}
        \caption{\textbf{Patterns underlying the statistical distributions of the responses of humans and machines.} Schematic illustration of the \textbf{a,} archetype I, along with \textbf{b,} three actual examples realizing it (i.e., the statistical distribution of the responses of physicists to S7, S10, and S27), and the  \textbf{c,}  archetype II, along with \textbf{d,} three actual examples realizing it (i.e., the statistical distribution of the responses of physicists to S4, S5, and S28).}
         \label{Figure-4}
\end{figure}

To better understand the capability of machines to emulate various sub-populations of humans, depending on their academic and research background, we examine the granularity of our results. In  Figure \ref{Figure-3}, we show the mean absolute differences in the mean agreement scores of several groups of participants, calculated using Equation \ref{delta} on the data displayed in Supplementary Figures 4, 5, and 6. We inspect sub-populations of physicists differing in research methodology (theoretical vs.\ experimental physics), scope (basic vs.\ applied physics), and level of experience (less vs.\ more than ten years), as well as for various areas of expertise. The mean absolute difference is quite insensitive to the specific sub-population of physicists, in that its variation exceeds the global value of $8.4\%$ only by 2.5\% at most. The largest variations are observed in the case of physicists with more than ten years of experience and those working in the field of nuclear and particle physics. An analogous conclusion is reached for philosophers of science, for which equivalent results are shown in Supplementary Figure 7.

\textcolor{black}{For physicists, we highlight that machines achieved agreement with the responses of humans at population level, even though the large language model was prompted only with minimal background attributes (e.g., disciplinary area and career stage), thus neglecting epistemic and theoretical stances as well as argumentative strategies that one would expect to play a key role in the development of philosophical views. The fact that the model could reproduce these distributions for physicists without direct philosophical cues and through minimal impersonation features is not trivial.}\footnote{\textcolor{black}{Since our framework is flexible, additional participant-level details could easily be incorporated into the prompts to test whether machines can capture perspectives when more granular data on survey respondents are available.}} \textcolor{black}{On the other hand,} \textcolor{black}{the resemblance between humans and machines observed for philosophers is perhaps less surprising, given that the prompts included their stated philosophical views. However, even in this case, the results are not entirely trivial, in that philosophers who shared the same philosophical position (e.g., structural realism) but differed in other attributes produced distinct outputs. This indicates that the responses of the model are sensitive to more than just the explicitly supplied philosophical labels.} \textcolor{black}{The modest differences in mean agreement scores may partly reflect limitations of the large language model, but they may equally arise from the survey design and the variability inherent in human responses. Similar to machines, human participants---even those sharing similar characteristics---often provide different scores for the same questions. Consequently, it is not expected that machine personas will reproduce individual human answers exactly. Instead, the ability to capture the overall distribution of responses represents a meaningful achievement.}

Despite these strong similarities in the mean agreement scores of populations of humans and machines, further analysis reveals significant differences in their statistical distributions, as detailed in Supplementary Figures 8 and 9 for physicists and philosophers of science, respectively. Specifically, we identify two main patterns underlying the vast majority of  these distributions. These patterns are described through the two distinct archetypes---referred to as archetypes I and II---depicted in Figure \ref{Figure-4} along with actual examples. In both archetypes, the responses of the population of machines exhibit a normal-like distribution of the agreement scores. However, humans manifest qualitatively different trends. In archetype I, the agreement scores of humans are uniformly distributed across the statements. In archetype II, the agreement scores are unevenly distributed, peaking at either end of the scale of the agreement score, thus resembling a skew-normal distribution. From a visual inspection of Supplementary Figures 8 and 9, we note that archetype II is approximately twice as recurrent as archetype I.

\begin{figure}[t]
    \centering
    \includegraphics[width=1\textwidth]{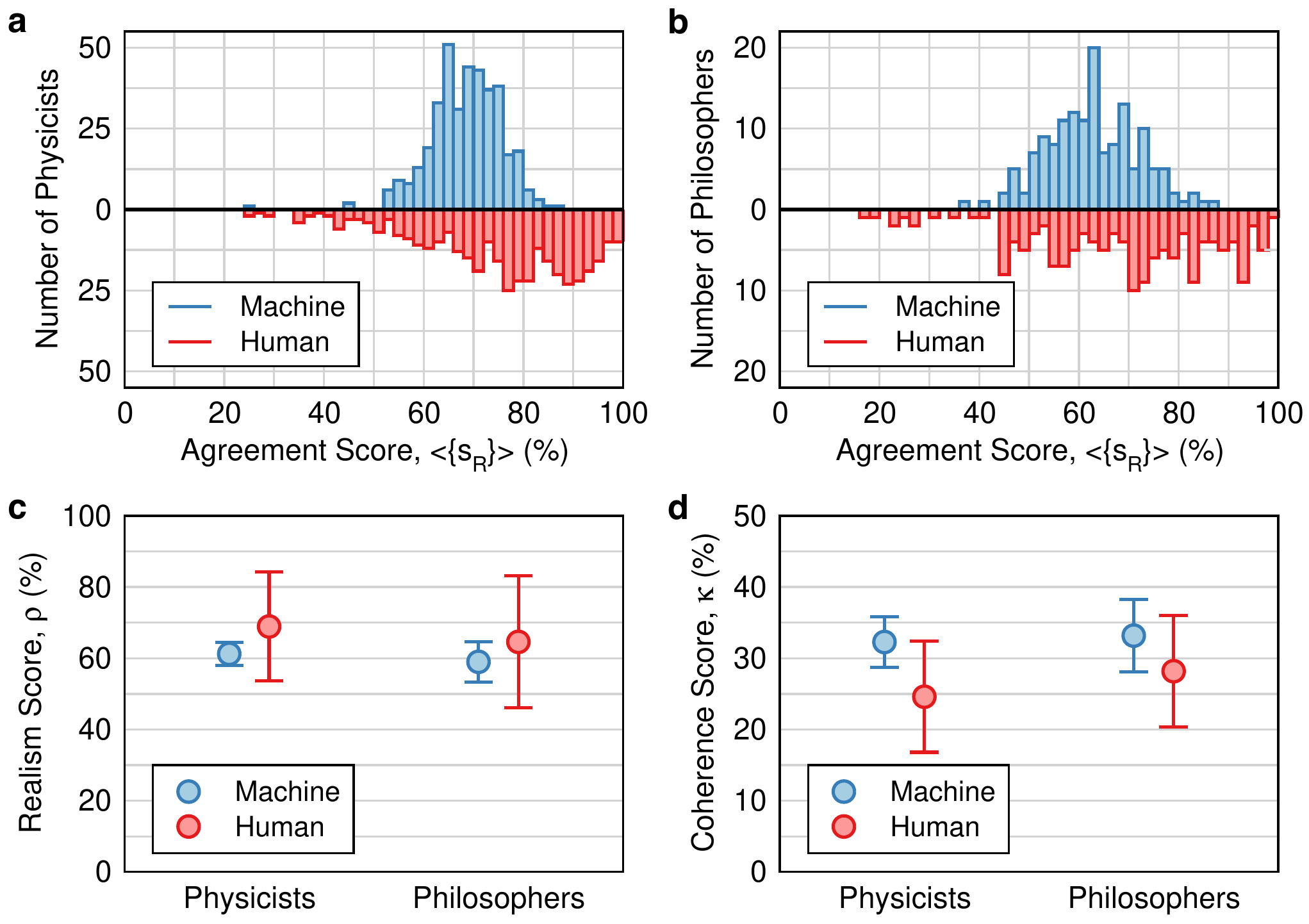}
    \caption{\textbf{Realist stance and internal coherence of humans and machines.} Statistical distribution of the individual agreement scores  assigned only to the realist statements ($<\{{s}\textsubscript{R}\}>$) listed in Table \ref{Table-1} by  \textbf{a,} physicists and  \textbf{b,} philosophers of science.   \textbf{c,} Realism score ($\rho$), as defined in Equation \ref{rho}, and \textbf{d,} coherence score ($\kappa$), as defined in Equation \ref{kappa}, for physicists and philosophers of science.}
         \label{Figure-5}
\end{figure}

To gain insights into the philosophical positions subscribed by human and machine populations, we focus on the two primary and opposing views in the realism debate, that is, scientific realism and instrumentalism. In Figure  \ref{Figure-5}(a,b), we compare the distribution of the individual mean agreement scores assigned by humans and machines to the realist statements listed in Table \ref{Table-1} (i.e., S1, S2, S4, S6, S8, S10, S11, S13, S14, S15, S17, S18, S20, S22) for both physicists and philosophers of science. These distributions are reminiscent of the archetypes displayed in Figure \ref{Figure-4}(a,c), with the population of machines exhibiting a normal-like distribution while the population of humans exhibiting an approximately uniform distribution, slightly skewed toward the highest values of agreement score. An analogous trend, albeit shifted to lower agreement scores, is observed when considering only the instrumentalist statements listed in Table \ref{Table-1}   (i.e., S3, S5, S7, S9, S12, S16, S23, S25), as shown in Supplementary Figure 10. The statistical distributions of the mean agreement scores assigned by several sub-populations of physicists to representative realist and instrumentalist statements are provided in Supplementary Figures 11 and 12, respectively.

To quantify the extent to which the populations of humans and machines favor scientific realism over instrumentalism, we determine the realism score, following the definition Henne and coworkers  \cite{Henne2024}, as  
\begin{equation}
\rho = <\{{s}\textsubscript{R}\}, - \{{s}\textsubscript{I}\} >,
\label{rho}
\end{equation}
where $\{ {s}\textsubscript{R} \}$ is the set of the agreement scores assigned selectively to the realist statements and $\{{s}\textsubscript{I} \}$ is the set of the agreement score assigned selectively to the instrumentalist statements listed in Table \ref{Table-1}. By construction, the realism score can range from 0\%, indicating a strict instrumentalist position, to 100\%, indicating a strict realist position. The realism scores pertaining to humans and machines are compared in Figure  \ref{Figure-5}(c) and listed in Supplementary Table 2.  Humans are more inclined toward realism than machines, with the realism score of the former being greater than the latter by  7.7\% for physicists and 5.6\% for philosophers of science. Importantly, we note that machines correctly reproduce the trend of human physicists being more realist than philosophers of science, although the difference in realism scores between physicists and philosophers is less pronounced for machines (for which it attains the value of 2.3\%) than humans (for which it attains the value of 4.3\%). We have confirmed that each pairwise distribution of realism scores is statistically distinct by verifying that the $p$-value is lower than 0.05. 

A key aspect in establishing a robust philosophical stance is the internal coherence, that is, the minimization---or, ideally, removal---of inherent contradictions within a given position. In the case of the realism-instrumentalism dichotomy, this translates to the assignment of agreement scores that are compatible across the statements describing the same philosophical view. Specifically, an internally coherent position would stem from high agreement scores when evaluating the realist (instrumentalist) statements and low agreement scores when evaluating the opposite instrumentalist (realist) statements. To quantify the internal coherence, we introduce a coherence score,
\begin{equation}
\kappa = \max[\sigma(\rho)] -  <\sigma(\rho)>,
 \label{kappa}
\end{equation}
where $\sigma(\rho)$ is the standard deviation pertaining to the agreement score associated with the statements included in the determination of the realism score (cf.\ Equation \ref{rho}) and $\max[\sigma(\rho)]$  is its maximum possible value, 50\%. The coherence score can range from 0\%, signaling a random distribution of the agreement scores across the statements and a consequent complete incoherence, to 50\%, signaling an absolute internal coherence within one of the two contrasting philosophical views. The coherence scores are shown in Figure \ref{Figure-5}(d) and listed in Supplementary Table 2. Notably, in their philosophical positions, machines are considerably more coherent than humans, with the difference in coherence scores between the former and the latter being 7.7\% for physicists and 5.0\% for philosophers of science. Importantly, machines are capable of replicating the trend of human philosophers being more internally coherent than physicists, although this difference is less pronounced for machines (0.9\%) compared to humans (3.6\%), as previously observed in the case of the realism score.

\section{Concluding remarks} In the spirit of the Turing test, we have developed a novel methodological framework to determine whether and to what degree a population of machines mirrors the philosophical views endorsed by a population of humans. Our methodology consists of three steps. 
First, a population of machines is instructed to impersonate a population of humans, emulating the background of each individual. 
Second, humans and machines are administered a questionnaire designed to survey various philosophical positions, with each participant rating their agreement with a series of statements. Third, the outcome of the survey is statistically analyzed to compare the views held by the  two populations.

Drawing from a recent study of Henne and coworkers  \cite{Henne2024}, we have employed this methodology in the case study of scientific realism, a long-standing philosophical debate seeking to understand if reality is as science describes it. We have examined the philosophical positions of a population of over 500  humans, comprising both physicists and philosophers of science, and the corresponding populations of machines, generated by means of a popular large language model via prompt engineering.  Our analysis has revealed that a population of machines endorses philosophical views that are, on average, very similar to those held by the corresponding population of humans, regardless of whether the respondents are physicists or philosophers of science, \textcolor{black}{even if the similarity at the \emph{individual} level can be limited, as shown in Supporting Figure 13.}
\textcolor{black}{Our work does \emph{not} constitute a genuine Turing test, as it did not involve a conversation between a human and a machine, nor the participation of a human judge.} However, the analogy of the philosophical judgments held by the populations of humans and machines---as corroborated by the invariably overlapping error bars of the metrics used to quantify them---implies that a hypothetical human judge is likely to fail to discern the nature of the two populations. Importantly, machines are able to reflect the nuances distinguishing the views of philosophers from those of physicists.

We have additionally observed that, as compared to humans, machines exhibit a weaker inclination toward scientific realism and a stronger coherence in their philosophical positions. Because the realism-instrumentalism debate is inherently underdetermined---in that no \emph{experimentum crucis} can be conceived to decisively adjudicate between these two opposed views---there is no single `true' interpretation of the relationship between science and reality that can serve as a benchmark to establish the exactness of the philosophical views held by humans or machines. However, if internal coherence is regarded as a measure of the strength of a given philosophical position, then one may provocatively suggest that machines are better philosophers than humans. \textcolor{black}{{The stronger coherence observed in machines is in line with earlier studies \cite{Long2025}, which have shown that large language models can provide more consistent answers across runs than humans do across individuals.}}

\textcolor{black}{We have verified the robustness of our framework by carrying out additional simulations employing either a different questionnaire or set of large language models. On the one hand, we have considered the survey of Beebe and Dellsén on scientific realism \cite{Beebe2020}. For physicists, we obtained a mean average difference between humans and machines of 7.2\%, in line with the corresponding value of 8.4\% determined for the questionnaire of Henne and coworkers \cite{Henne2024}. On the other hand, we tested multiple versions of GPT models by applying them to the survey of Henne and coworkers \cite{Henne2024}. For physicists, we obtained comparable mean average differences of 8.4\%, 8.6\%, and 8.5\% for GPT-3.5-turbo, GPT-4, and GPT-5-nano, respectively, despite notable differences in features such as parameter count, context window length, reasoning ability, and domain-specific capabilities. These values demonstrate that our results are robust across different questionnaires and large language models.}
\textcolor{black}{Unlike prior works  \cite{Zhao2025, Simmons2024}, in our approach we do not provide the large language model with illustrative examples from which they could learn and generalize. Instead, we directly probe whether the model possesses intrinsic knowledge of complex topics---such as the philosophical perspectives of physicists---and whether they can reproduce nuanced, non-trivial response patterns solely based on their training. This design enables us to test, in a principled way, whether large language models can mirror the diversity and structure of real human populations without additional fine-tuning.}

\textcolor{black}{To conclude, our methodology can be readily applied to other surveys across the philosophy of science, such as those assessing the views of a wide variety of scholars on, e.g., theoretical virtues \cite{Schindler:2022}, values in science \cite{Steel:2017}, and scientific practices \cite{Robinson:2019}, as well as on broader issues relevant to other philosophical domains \cite{Bourget:2014, Bourget:2023}, such as philosophical intuition \cite{Kuntz:2011} and folk psychology \cite{Hewson1994}.} Given the close resemblance between human and machine populations in their average responses to philosophical questions, our methodological framework may open a new avenue to advance research in the \textcolor{black}{empirical social sciences} by deploying a population of machines in lieu of a target population of humans,\footnote{\textcolor{black}{One may also wonder whether large language models would be able to swiftly track shifts in response to changes occurring in the population impersonated (e.g., the community of physicists). Broadly, large language models are primarily retrospective: their outputs reflect the distribution of views present in their training corpus. Consequently, their ability to capture rapid shifts in the perspective of a population depends on how quickly such changes are incorporated into updated training sets. However, this limitation may be mitigated through emerging strategies, such as retrieval-augmented generation and domain-specific fine-tuning, which aim to reduce the lag between new developments and model outputs. Moreover, recent studies have explored the possibility that large language models can generate plausible projections or reason about trends based on historical data \cite{Nako2025}. Although this forward-looking reasoning is constrained by the quality and representativeness of the input data and prompts, these approaches lay the groundwork for potentially predicting shifts in the perspectives of a target population.}} possibly accelerating the administration of otherwise tedious survey-based studies and mitigating the reproducibility issues plaguing them \cite{Cova:2021}.


\bibliographystyle{apalike}   
\bibliography{Final-Bibliography} 

 \end{document}